\documentclass{article}
\usepackage{graphicx}
\usepackage{amsmath}
\usepackage{amsfonts}
\usepackage{amssymb}
\newtheorem{theorem}{Theorem}

\newtheorem{example}[theorem]{Example}

\newtheorem{proposition}[theorem]{Proposition}
\newtheorem{remark}[theorem]{Remark}

\newenvironment{proof}[1][Proof]{\textbf{#1.} }{\ \rule{0.5em}{0.5em}}

\begin{document}

\title{A quaternionic generalisation of the Riccati differential equation}
\author{Viktor Kravchenko,\\{\small Univ. do Algarve, Dep. Math., Campus de Gambelas, 8000 Faro, Portugal}
\and Vladislav Kravchenko,\\{\small Esc. Sup. de Ing. Mec. y El\'{e}c. del Inst. Polit. Nac., Dept. de Telecom.,}\\{\small Unidad Zacatenco, C.P. 07738, D.F., M\'{e}xico}
\and Benjamin Williams\\{\small Esc. Sup. de Fis. Mat. del Inst. Polit. Nac., }\\{\small Unidad Zacatenco, C.P. 07738, D.F., M\'{e}xico}}
\maketitle
\begin{abstract}
A quaternionic partial differential equation is shown to be a generalisation
of the traditional Riccati equation and its relationship with the
Schr\"{o}dinger equation is established. Various approaches to the problem of
finding particular solutions to this equation are explored, and the
generalisations of two theorems of Euler on the Riccati equation, which
correspond to this partial differential equation, are stated and proved.
\end{abstract}

\section{Introduction}

The Riccati equation
\[
\partial u=pu^{2}+qu+r,
\]
where $p,\ q$ and $r$ are functions, has received a great deal of attention
since a particular version was first studied by Count Riccati in 1724, owing
to both its peculiar properties and the wide range of applications in which it
appears. \ For a survey of the history and classical results on this equation,
see for example \cite{wats}, \cite{davis} and \cite{reid}. \ This equation can
be reduced to its canonical form \cite{bogd},
\begin{equation}
\partial y+y^{2}=-v,\label{ricc1}%
\end{equation}
and this is the form that we will consider.

One of the reasons for which the Riccati equation has so many applications is
that it is related to the general second order homogeneous differential
equation. \ In particular, the one-dimensional Schr\"{o}dinger equation
\begin{equation}
-\partial^{2}u-vu=0\label{schrod1}%
\end{equation}
where $v$ is a function, is related to the (\ref{ricc1}) by the easily
inverted substitution $y=\frac{\partial u}{u}.$ \ This substitution, which as
its most spectacular application reduces Burger's equation to the standard
one-dimensional heat equation, is the basis of the well-developed theory of
logarithmic derivatives for the integration of nonlinear differential
equations \cite{march}. \ A generalisation of this substitution will be used
in this work.

A second relation between the one-dimensional Schr\"{o}dinger equation and the
Riccati equation is as follows. \ The one-dimensional Schr\"{o}dinger operator
can be factorised in the form
\[
-\partial^{2}-v(x)=-(\partial+y(x))(\partial-y(x))
\]
if and only if (\ref{ricc1}) holds.

\bigskip Among the peculiar properties of the Riccati equation stand out two
theorems of Euler, dating from 1760. \ The first of these \cite{wats} states
that if a particular solution $y_{0}$ of the Riccati equatuion is known, the
substitution $y=y_{0}+z$ reduces (\ref{ricc1}) to a Bernoulli equation which
in turn is reduced by the substitution $z=\frac{1}{u}$ to a first order linear
equation. \ Thus given a particular solution of the Riccati equation, the
general solution can be found in two integrations. \ The second of these
theorems \cite{reid} states that given two particular solutions $y_{0}%
,\ y_{1}$ of the Riccati equation, the general solution can be found in the
form
\begin{equation}
y=\frac{ky_{0}\exp(\int y_{0}-y_{1})-y_{1}}{k\exp(\int y_{0}-y_{1}%
)-1}\label{2sol}%
\end{equation}
where $k$ is a constant. \ That is, given two particular solutions of
(\ref{ricc1}), the general solution can be found in one integration.

Other interesting properties are those discovered by Picard and Weyr
\cite{wats}. \ The first is that given a third particular solution $y_{3}$,
the general solution can be found without integrating. \ That is, an explicit
combination of three particular solutions gives the general solution. The
second is that given a fourth particular solutions $y_{4}$, the cross ratio
\[
\frac{(y_{1}-y_{2})(y_{3}-y_{4})}{(y_{1}-y_{4})(y_{3}-y_{2})}%
\]
is a constant.

This article is concerned with a quaternionic generalisation of the Riccati
equation and versions of the above-mentioned theorems of Euler corresponding
to this generalisation. \ Some necessary notation will be introduced in
Section \ref{quats}. \ In Section \ref{3dr} we propose the quaternionic
generalisation of the Riccati equation, which is shown to be a good
generalisation for various reasons, including that it is related to the
three-dimensional Schr\"{o}dinger equation
\begin{equation}
\triangle u+vu=0\label{schrod}%
\end{equation}
(where $\triangle$ is the three-dimensional laplacian) in the same ways as the
Riccati equation is related to (\ref{schrod1}). \ In Section \ref{part}, we
turn our attention to cases in which particular solutions of (\ref{ricc3d})
can be found, some of which differ considerably from the one-dimensional case.
\ Finally, in Section \ref{3da}, generalisations of \ Euler's theorems will be
stated and proved. 

\section{Preliminaries}

The complex numbers and complex quaterions are denoted by $\mathbb{C}%
,\ \mathbb{H(C)}$ respectively. The latter consists of elements of the form%

\[
a=\sum\limits_{k=0}^{3}a_{k}i_{k}%
\]
where the $a_{k}\in\mathbb{C}$ and the base units $i_{k}$ satisfy the
following rules of multiplication%

\begin{align*}
i_{0}^{2}  &  =i_{0}=-i_{k}^{2},\ i_{0}i_{k}=i_{0}i_{k}=i_{k},\ k=1,2,3,\\
i_{1}i_{2}  &  =-i_{2}i_{1}=i_{3},\ i_{2}i_{3}=-i_{3}i_{2}=i_{1},\ i_{3}%
i_{1}=-i_{1}i_{3}=i_{2}.
\end{align*}
The complex unit $i$ commutes with the $i_{k}.$ \ Frequently it is useful to
consider a quaternion $a$ as being the sum of a scalar and a vector part,
denoted respectively%

\[
a_{0}:=\operatorname*{Sc}(a),\ \overrightarrow{a}:=\operatorname*{Vec}%
(a)=\sum\limits_{k=1}^{3}a_{k}i_{k}.
\]
Conjugation is defined as follows,
\[
\overline{a}:=a_{0}-\overrightarrow{a}%
\]
and the modulus is
\[
\left|  a\right|  ^{2}=a\cdot\overline{a}=a_{0}^{2}+a_{1}^{2}+a_{2}^{2}%
+a_{3}^{2}%
\]
so that in particular $\overrightarrow{a}^{2}=-\left|  \overrightarrow
{a}\right|  ^{2}.$

Note that in terms of scalars and vectors, the quaternionic product can be
written
\[
ab=(a_{0}+\overrightarrow{a})(b_{0}+\overrightarrow{b})=a_{0}b_{0}%
+a_{0}\overrightarrow{b}+b_{0}\overrightarrow{a}-\left\langle \overrightarrow
{a},\overrightarrow{b}\right\rangle +\left[  \overrightarrow{a}\times
\overrightarrow{b}\right]
\]
where $\left\langle a,b\right\rangle $ is the standard inner product and
$\left[  a\times b\right]  $ the standard cross product in $\mathbb{R}^{3}.$
\ In particular%

\begin{equation}
\{\overrightarrow{a},\overrightarrow{b}\}=-2\left\langle \overrightarrow
{a},\overrightarrow{b}\right\rangle \label{antisum}%
\end{equation}
where $\{a,b\}=ab+ba$ is the standard anticommutator.

In what follows functions $g:\mathbb{\Omega}\rightarrow\ \mathbb{H(C)}$ will
be considered, where $\Omega$ is some domain in $\mathbb{R}^{3}$. \ The
Moisil-Theodoresco operator $D$ is defined on differentiable functions $g$ as follows:%

\[
Dg=\sum\limits_{k=1}^{3}i_{k}\partial_{k}g
\]
where $\partial_{k}=\frac{\partial}{\partial x_{k}}.$ Due to properties of the
quaternionic product, this can be written
\[
Dg=-\operatorname*{div}\overrightarrow{g}+\operatorname*{grad}g_{0}%
+\operatorname*{rot}\overrightarrow{g}.
\]
Thus it follows that%

\begin{align*}
\operatorname*{Sc}(D\overrightarrow{g})  &  =-\operatorname*{div}%
\overrightarrow{g},\\
\operatorname*{Vec}(D\overrightarrow{g})  &  =\operatorname*{rot}%
\overrightarrow{g},
\end{align*}
and for scalar functions $u$%
\[
Du=\operatorname*{grad}u.
\]

\bigskip The theorem of Leibnitz for this operator is the following: given a
differentiable scalar function $u$ and a differentiable quaternionic function
$g,$%
\begin{equation}
D(ug)=D(u)g+uD(g).\label{chain}%
\end{equation}

The logarithmic derivative (in the sense of Marchenko) of a scalar function
$u$ such that $u\neq0$ in $\Omega,$ is defined as
\[
\check{\partial}u=u^{-1}Du.
\]
The function $\check{\partial}u$ is a vector. \ The derivative is logarithmic
in the following sense: given two scalar functions $u_{1},u_{2}$ that do not
vanish in $\Omega$, formula (\ref{chain}) implies that
\[
\check{\partial}(u_{1}u_{2})=\check{\partial}u_{1}+\check{\partial}u_{2}.
\]

\section{\bigskip A three-dimensional generalisation of the Riccati
equation\label{3dr}}

The substitution of the Jackiw-Nohl-Rebbi-'t\ Hooft ansatz \cite{jackiw} in
the self-duality equation can be written \cite{gursey} in the following
quaternionic form%

\begin{equation}
\partial_{t}g+Dg+\left|  g\right|  ^{2}=0\label{rhubarb}%
\end{equation}
where the subscript $t$ denotes differentiation with respect to time. \ This
equation has obvious formal similarities with equation (\ref{ricc1}). \ In
\cite{krav2} the relation of (\ref{rhubarb}) to the F\"{u}ter operator
\[
\partial_{t}+D
\]
was shown. In particular, it was shown that for any $f\in\ker(\partial
_{t}+D),$ the function
\[
\frac{2(\operatorname*{grad}f_{0}-\operatorname*{div}\overrightarrow{f}%
)}{f_{0}}%
\]
is a solution of (\ref{rhubarb}), that is, a class of instantons was obtained.
\ In what follows we will concentrate on the case of time independent, purely
vectorial quaternionic functions, but the nonhomogeneous equation will be considered.

\bigskip The following result generalises to three dimensions the relation,
mentioned in the introduction, between the one-dimensional Schr\"{o}dinger
operator and the Riccati differential equation via the logarithmic derivative.

\begin{proposition}
$\varphi$ is a solution of (\ref{schrod}) if and only if $\overrightarrow
{f}:=\check{\partial}\varphi$ is a solution of
\begin{equation}
D\overrightarrow{f}+\overrightarrow{f}^{2}=v\label{ricc3d}%
\end{equation}
\end{proposition}

\begin{proof}
\bigskip Suppose that there exists a function $\varphi$ such that
$\overrightarrow{f}=\check{\partial}\varphi.$ \ Applying (\ref{chain}) gives
\begin{align*}
D\overrightarrow{f} &  =\frac{1}{\varphi^{2}}\left\langle \nabla\varphi
,\nabla\varphi\right\rangle -\frac{1}{\varphi}\triangle\varphi,\\
\overrightarrow{f}^{2} &  =-\frac{1}{\varphi^{2}}\left\langle \nabla
\varphi,\nabla\varphi\right\rangle ,
\end{align*}
so that $-\frac{1}{\varphi}\triangle\varphi=v,$ or equivalently $\triangle
\varphi+v\varphi=0.$ Conversely, given a solution $\varphi$ of (\ref{schrod}),
$\overrightarrow{f}=\check{\partial}\varphi$ is a solution of (\ref{ricc3d}).
\end{proof}

\bigskip

In \cite{Swansolo} and \cite{Swan} it was shown that the three-dimensional
Schr\"{o}dinger operator can be factorised in the following way
\[
-\triangle-vI=(D+M^{\overrightarrow{f}})(D-M^{\overrightarrow{f}})
\]
where $I$ is the identity operator and $M^{\overrightarrow{f}}%
q:=q\overrightarrow{f},$ if and only if equation (\ref{ricc3d}) holds.

Thus the two relations between the Riccati equation (\ref{ricc1}) and the
one-dimensional Schr\"{o}dinger equation (\ref{schrod1}) mentioned in the
introduction have natural counterparts relating (\ref{ricc3d}) and the
three-dimensional Schr\"{o}dinger equation (\ref{schrod}). \ These
relationships suggest that (\ref{ricc3d}) can be considered a good
generalisation of (\ref{ricc1}). \ It should also be noted that if
$\overrightarrow{f}=f_{k}(x_{k})i_{k},$ then (\ref{ricc3d}) is reduced to
\[
\partial_{k}f_{k}+f_{k}^{2}=-v.
\]
That is, a one-dimensional solution of (\ref{ricc3d}) is a solution of
(\ref{ricc1}). \ Equation (\ref{ricc3d}) will be refered to as the Riccati PDE.

Note that the scalar and vector components of equation (\ref{ricc3d}) are
respectively
\begin{align*}
-\operatorname*{div}\overrightarrow{f}+\overrightarrow{f}^{2} &  =v,\\
\operatorname*{rot}\overrightarrow{f} &  =0.
\end{align*}
The second equation implies that for a simply-connected domain $\Omega,$ there
exists a scalar function $\varphi$ such that $\overrightarrow{f}%
=\operatorname*{grad}\varphi.$ \ Substituting this in the first equation
gives
\begin{equation}
\triangle\varphi+\left\langle \nabla\varphi,\nabla\varphi\right\rangle
=-v.\label{equiv}%
\end{equation}
This equivalence of the Riccati PDE with a scalar elliptic partial
differential equation will be used frequently in what follows.

It should also be noted that if the function $v$ in (\ref{ricc3d}) is zero,
the substitution $\overrightarrow{f}=\check{\partial}\varphi$ reduces the
equation to
\[
\triangle\varphi=0.
\]
Thus the homogeneous Riccati equation can be solved explicitly, and its
solutions are of the form $\check{\partial}\varphi$, where $\varphi\in
\ker\triangle.$ \ \ This generalises the highly restricted class of solutions
of the homogeneous equation (\ref{ricc1}), which are precisely functions of
the form
\[
\frac{1}{x+c},
\]
$c$ constant.

\bigskip

\section{\bigskip Particular solutions of the equation\label{part}}

\bigskip In the next section it will be shown that given one solution of the
Riccati PDE\ it can be linearised. \ Thus in this section we discuss some
possibilities for obtaining particular solutions of (\ref{ricc3d}). \ First we
note that if the function $v$ is of the form
\[
v(x)=v_{1}(x_{1})+v_{2}(x_{2})+v_{3}(x_{3}),
\]
then assuming that $\overrightarrow{f}=f_{1}(x_{1})i_{1}+f_{2}(x_{2}%
)i_{2}+f_{3}(x_{3})i_{3},$ equation (\ref{ricc3d}) reduces to the system of
Riccati ordinary differential equations
\[
\partial_{k}f_{k}+f_{k}^{2}=-v_{k},\ k=1,2,3.
\]
Thus in this case, a particular solution of (\ref{ricc3d}) can be found if and
only if each of the above equations can be solved. \ Obviously, if any of the
$v_{k}$'s are zero, this task is greatly simplified. \ This situation
corresponds to the Schr\"{o}dinger equation (\ref{schrod}) with potential $v$
of the form given above, in which case the variables can be separated.

The existence of a large class of solutions of the homogeneous equation, as
described in Section \ref{3dr}, motivates the following procedure, which
reduces (\ref{ricc3d}) to various scalar differential equations. Substituting
the sum of two functions $\overrightarrow{f}_{1},\ \overrightarrow{f}_{2}\in
C^{1}(\Omega)$ into equation (\ref{ricc3d}) gives the following possible
decomposition:
\begin{align*}
D\overrightarrow{f}_{1}+\overrightarrow{f}_{1}^{2} &  =v_{1},\\
D\overrightarrow{f}_{2}+\overrightarrow{f}_{2}^{2} &  =v_{2},\\
\{\overrightarrow{f}_{1},\overrightarrow{f}_{2}\} &  =v_{3},\\
v &  =v_{1}+v_{2}+v_{3.}%
\end{align*}

In particular, if $\overrightarrow{f}_{1},\ \overrightarrow{f}_{2}$ are
solutions of the homogeneous equation, this is reduced to
\begin{equation}
\{\overrightarrow{f}_{1},\overrightarrow{f}_{2}\}=-2\left\langle
\check{\partial}\varphi_{1},\check{\partial}\varphi_{2}\right\rangle
=v,\label{logs}%
\end{equation}
where $\varphi_{1},\ \varphi_{2}$ are harmonic functions.

\bigskip If $\varphi_{1}=x_{1},\ \overrightarrow{f}_{1}=\frac{i_{1}}{x_{1}},$
this becomes
\[
\partial_{1}\varphi_{2}=-\frac{1}{2}vx_{1}\varphi_{2}%
\]
which has solution
\[
\varphi_{2}=A(x_{2},x_{3})\exp(-\frac{1}{2}\int vx_{1}dx_{1})
\]
where $A(x_{2},x_{3})$ is an arbitrary function. Thus if $\varphi_{2}\in
\ker\triangle$ and $\varphi_{2}$ satisfies the above equation, the sum
\[
\frac{i_{1}}{x_{1}}+\check{\partial}\varphi_{2}%
\]
is a particular solution of the Riccati PDE.

If instead of choosing $\varphi_{1}$ as above $\varphi_{1}=\varphi_{2}$ is
substituted in (\ref{logs}), the eikonal equation
\begin{equation}
2(\check{\partial}\varphi)^{2}=-v\label{eik}%
\end{equation}
results. \ Thus for a scalar function $\varphi\in\ker\triangle_{3},$ which is
also a solution of the above eikonal equation, $\overrightarrow{f}%
=2\check{\partial}\varphi$ is a solution of (\ref{ricc3d}).

\begin{example}
\bigskip Let $\varphi$ be the fundamental solution of the laplacian
$\triangle$,
\[
\varphi=\frac{1}{4\pi\left|  x\right|  }.
\]
This function is harmonic and positive in any domain $\Omega$ which does not
include the origin. \ Furthermore it satisfies equation (\ref{eik}) with
\[
v=\frac{1}{\left|  x\right|  ^{2}}.
\]
Thus
\[
\overrightarrow{f}=\frac{-2\overrightarrow{x}}{\left|  x\right|  ^{2}}%
\]
is a solution of (\ref{ricc3d}).
\end{example}

\bigskip

\section{\bigskip\bigskip Generalisations of Euler's theorems on the Riccati
equation\label{3da}}

\bigskip We now state and prove the generalisations to the Riccati PDE of
Euler's theorems on the Riccati equation that were mentioned in the
introduction. \ The first of these theorems states that given a particular
solution of the Riccati differential equation, the equation can be linearised. \ 

\begin{proposition}
(Generalisation of Euler's first theorem)\bigskip\bigskip\ Let
$\overrightarrow{h}=\operatorname*{grad}\xi$ be an arbitrary particular
solution of (\ref{ricc3d}). \ Then
\begin{equation}
\overrightarrow{f}=\overrightarrow{g}+\overrightarrow{h}\label{sum}%
\end{equation}
is also a solution of (\ref{ricc3d}), where $\overrightarrow{g}=\check
{\partial}\Psi$ and $\Psi$ is a solution of the equation
\begin{equation}
\triangle\Psi+2\left\langle \nabla\xi,\nabla\Psi\right\rangle =0,\label{trans}%
\end{equation}
or equivalently of
\begin{equation}
\operatorname*{div}(e^{2\xi}\nabla\Psi)=0.\label{div}%
\end{equation}
\end{proposition}

\begin{proof}
Substituting (\ref{sum}) in (\ref{ricc3d}) gives
\[
D\overrightarrow{g}+\{\overrightarrow{h,}\overrightarrow{g}\}+\overrightarrow
{g}^{2}=0
\]
or alternatively, using (\ref{antisum})
\begin{equation}
D\overrightarrow{g}-2\left\langle \overrightarrow{h},\overrightarrow
{g}\right\rangle +\overrightarrow{g}^{2}=0.\label{ours}%
\end{equation}
Note that, as in (\ref{ricc3d}), the vector part of (\ref{ours}) is
$\operatorname*{rot}\overrightarrow{g}=0,$ so that
\[
\overrightarrow{g}=\operatorname*{grad}\Phi
\]
for some function $\Phi.$ \ If $\Psi=e^{\Phi},$ this is equivalent to
\[
\overrightarrow{g}=\check{\partial}\Psi.
\]
Equation (\ref{ours}), written in terms of $\Psi,$ is
\[
-\frac{1}{\Psi^{2}}(\nabla\Psi)^{2}-\frac{1}{\Psi}\triangle\Psi-\frac{2}{\Psi
}\left\langle \nabla\xi,\nabla\Psi\right\rangle +\frac{1}{\Psi^{2}}(\nabla
\Psi)^{2}=0,
\]
so that (\ref{ours}) is equivalent to
\[
\triangle\Psi+2\left\langle \nabla\xi,\nabla\Psi\right\rangle =0.
\]
Noting that
\[
\operatorname*{div}(e^{2\xi}\nabla\Psi)=2e^{2\xi}\left\langle \nabla\xi
,\nabla\Psi\right\rangle +e^{2\xi}\triangle\Psi=e^{2\xi}(\triangle
\Psi+2\left\langle \nabla\xi,\nabla\Psi\right\rangle ),
\]
this equation can be rewritten in the form
\[
\operatorname*{div}(e^{2\xi}\nabla\Psi)=0.
\]
\end{proof}

\bigskip Equation (\ref{trans}) is the well-known transport equation, which
appears for example in the Ray Method of approximations of solutions to the
wave equation, coupled with the eikonal equation\cite{babs}. \ Equation
(\ref{div}) appears in various applications, for example in electrostatics
\cite{landau}, where $e^{2\xi}$ is the dielectric permeability and $\Psi$ is
the electric field potential, and as the continuity equation of hydromechanics
in the case of a steady flow, where $e^{2\xi}$ is the density of the medium
\cite{Fox}.

\begin{remark}
\bigskip From (\ref{div}) we have that
\[
e^{2\xi}\nabla\Psi=\operatorname*{rot}\overrightarrow{s}%
\]
for some vector-valued function $\overrightarrow{s},$ or
\[
\nabla\Psi=e^{-2\xi}\operatorname*{rot}\overrightarrow{s},
\]
where $\overrightarrow{s}$ must satisfy the condition
\[
\operatorname*{rot}(e^{-2\xi}\operatorname*{rot}\overrightarrow{s})=0
\]
as $e^{-2\xi}\operatorname*{rot}\overrightarrow{s}$ must be the gradient of
some function.
\end{remark}

As mentioned in the introduction, \bigskip given two particular solutions
$y_{1},\ y_{2\text{ }}$of the Riccati ordinary differential equation the
general solution can be found in one integration. \ A natural question is
whether this property extends to the Riccati PDE. \ The form (\ref{2sol}) of
the general solution found in this case, suggests a similar substitution in
(\ref{ricc3d}). \ This line of reasoning gives the following result.

\begin{proposition}
(Generalisation of Euler's second theorem)\bigskip\ Let $\overrightarrow
{h}_{1}=\operatorname*{grad}\xi_{1}$ and $\overrightarrow{h}_{2}%
=\operatorname*{grad}\xi_{2}$ be two particular solutions of \ (\ref{ricc3d}).
\ Then there exists a scalar function $\varphi$ such that
\[
\overrightarrow{f}=\nabla\varphi=\frac{\overrightarrow{h}_{1}w-\overrightarrow
{h}_{2}}{w-1}%
\]
is also a solution of (\ref{ricc3d}), where $w=Ae^{\xi_{1}-\xi_{2}}%
,\ A\in\mathbb{C}.$
\end{proposition}

\begin{proof}
\bigskip Equation (\ref{ricc3d}) is equivalent to%

\begin{equation}
\triangle\varphi+\left\langle \nabla\varphi,\nabla\varphi\right\rangle
=-v\label{rhu2}%
\end{equation}
where $\overrightarrow{f}=\operatorname*{grad}\varphi.$ \ The scalar functions
$\xi_{1}$ and $\xi_{2}$ are two solutions of (\ref{rhu2}). \ Substituting the
expression
\begin{equation}
\frac{\nabla\xi_{1}w-\nabla\xi_{2}}{w-1}\label{2solns}%
\end{equation}
for $\nabla\varphi$ into this equation, where $w$ is a scalar function, gives
\begin{align*}
\triangle\varphi &  =\operatorname*{div}\left(  \frac{\nabla\xi_{1}w-\nabla
\xi_{2}}{w-1}\right)  \\
&  =\frac{(w-1)(w\triangle\xi_{1}+\nabla w\cdot\nabla\xi_{1}-\triangle\xi
_{2})-\nabla w\cdot(w\nabla\xi_{1}-\nabla\xi_{2})}{(w-1)^{2}}%
\end{align*}
and
\[
\left\langle \nabla\varphi,\nabla\varphi\right\rangle =\frac{(w\nabla\xi
_{1})^{2}-2w\nabla\xi_{1}\cdot\nabla\xi_{2}+(\nabla\xi_{2})^{2}}{(w-1)^{2}}.
\]
Simplifying and using the fact that $\xi_{1}$ and $\xi_{2\text{ }}$are
solutions of (\ref{rhu2}), the equation is reduced to
\[
\nabla\log w=\frac{\nabla w}{w}=\nabla(\xi_{1}-\xi_{2}),
\]
so that
\[
w=Ae^{\xi_{1}-\xi_{2}}%
\]
for an arbitrary constant $A\in\mathbb{C}.$ \ It remains to show that the
expression (\ref{2solns}) is the gradient of some scalar function $\varphi.$
\ This is the case if the rotational of (\ref{2solns}) disappears. \ This is
shown as follows, where the identities $\operatorname*{rot}\nabla
\varphi=0,\ \overrightarrow{f}\times\overrightarrow{f}=0$ are used.
\begin{align*}
\operatorname*{rot}\left(  \frac{\nabla\xi_{1}w-\nabla\xi_{2}}{w-1}\right)
&  =\operatorname*{rot}\left(  \frac{\nabla\xi_{1}w}{w-1}\right)
-\operatorname*{rot}\left(  \frac{\nabla\xi_{2}}{w-1}\right)  \\
&  =\operatorname*{grad}\frac{w}{w-1}\times\nabla\xi_{1}+\frac{w}%
{w-1}\operatorname*{rot}\nabla\xi_{1}\\
&  -\operatorname*{grad}\frac{1}{w-1}\times\nabla\xi_{2}-\frac{1}%
{w-1}\operatorname*{rot}\nabla\xi_{2}\\
&  =\operatorname*{grad}\frac{w}{w-1}\times\nabla\xi_{1}-\operatorname*{grad}%
\frac{1}{w-1}\times\nabla\xi_{2}\\
&  =\frac{(w-1)\nabla w-w\nabla w}{(w-1)^{2}}\times\nabla\xi_{1}+\frac{\nabla
w}{(w-1)^{2}}\times\nabla\xi_{2}\\
&  =\frac{\nabla w}{(w-1)^{2}}\times\nabla(\xi_{2}-\xi_{1})\\
&  =-\frac{Ae^{\xi_{1}-\xi_{2}}}{(w-1)^{2}}\nabla(\xi_{2}-\xi_{1})\times
\nabla(\xi_{2}-\xi_{1})\\
&  =0.
\end{align*}
\end{proof}

\bigskip\ It must be noted that the new solution gained is not necessarily the
general solution, but a larger class of solutions. \ The following example
illustrates this.

\begin{example}
Consider (\ref{ricc3d}) with $v=-1,$%
\[
D\overrightarrow{f}+\overrightarrow{f}^{2}=-1.
\]
Two solutions of this equation are $h_{1}=i_{1}=\operatorname*{grad}%
x_{1},\ h_{2}=i_{2}=\operatorname*{grad}x_{2}.$ \ Applying the above result
gives the class of solutions
\[
\frac{i_{1}Ae^{x_{1}-x_{2}}-i_{2}}{Ae^{x_{1}-x_{2}}-1},\ A\in\mathbb{C},
\]
\ however a third solution $h_{3}=i_{3}$ is not included in the above
expression as a special case.
\end{example}

\end{document}